\documentclass[aps,pre,showpacs,preprint,floatfix]{revtex4-1}
\usepackage{graphicx}

\begin{document}

%% \title{Propagation of diseases through a mobile agent system}
\title{Reaction processes among self-propelled particles}

\author{Fernando Peruani}
\affiliation{Universit{\'e} C{\^o}te d' Azur, Laboratoire J.A. Dieudonn{\'e},  UMR CNRS 7351, Parc Valrose, F-06108 Nice Cedex 02, France} 

\author{Gustavo Sibona}
\affiliation{CONICET and Facultad de Matem\'atica, Astronom\'ia y F\'isica, 
Universidad Nacional de C\'ordoba, C\'ordoba, Argentina}

\begin{abstract}
We study a system of self-propelled disks that perform run-and-tumble motion, where particles can adopt more than one internal state. One of those internal states can be transmitted to another particle if the particle carrying this state maintains physical contact with another particle for a finite period of time. We refer to this process as a {\it reaction process} and to the different internal states as {\it particle species} making an analogy to chemical reactions. The studied system may fall into an absorbing phase, where due to the disappearance of one of the particle species no further reaction can occur or remain in an active phase where particles constantly react. Combining individual-based simulations and mean-field arguments, we study the dependence of the equilibrium densities of particle species with motility parameters, specifically the active speed $v_0$ and tumbling frequency $\lambda$. We find that the equilibrium densities of particle species exhibit two very distinct, non-trivial scaling regimes with $v_0$ and $\lambda$ depending on whether the system is in the so-called ballistic or diffusive regime. Our mean-field estimates lead to an effective renormalization of reaction rates that allow building the phase-diagram $v_0$--$\lambda$ that separates the absorbing and active phase. We find an excellent agreement between numerical simulations and estimates. This study is a necessary step to an understanding of phase transitions into an absorbing state in active systems and sheds light on the spreading of information/signaling among moving elements.
\end{abstract}

\maketitle

%%%%%%%%%% TEXT %%%%%%%%%%%%%%%%%

\section{Introduction}
Systems of self-propelled particles  are found in biology across scales, from  bacteria~\cite{zhang2010,  cisneros2011, peruani2012} to animal groups~\cite{ballerini2008,ginelli2015}. There exist also man-made self-propelled systems such as chemically driven particles~\cite{howse2007, uspal2015}, vibration-driven disks~\cite{deseigne2008} and various types of active rollers~\cite{bricard2015,kaiser2017flocking} among  many other examples.  
Requiring particles to convert energy into work to self-propel in dissipative media, 
self-propelled particle systems are intrinsically nonequilibrium systems~\cite{vicsek2012, marchetti2013,bechinger2016}.  
Given the nonequilibrium nature of these systems, self-propelled particle systems display a large variety of phenomena that cannot be found in equilibrium systems 
as for instance the spontaneous, self-organized emergence of long-range order in the form of large-scale collective motion  in two-dimensions that was initially found in models~\cite{vicsek1995novel,toner1995long,ginelli2010large} 
and later confirmed to exist in real-world systems~\cite{deseigne2008,bricard2015,nishiguchi2017long}.  
It is worth noting that it was recently observed that 
the presence of spatial heterogeneities such as imperfections on the substrate where particles move -- typically present in real systems --  prevents 
the emergence of large-scale collective motion~\cite{chepizhko2013a, peruani2018}. 
Not surprisingly,  most experimental self-propelled systems 
do not display global collective motion and particle motion remains diffusive at large time scales~\cite{zhang2010, cisneros2011, peruani2012, howse2007}.  
However, even in the absence of global order, the self-propulsion of active particles induces remarkable nonequilibrium features such as 
non-equilibrium clustering~\cite{ramaswamy2017,peruani2006, yang2010, peruani2012}, nonequilibrium phase-separation~\cite{fily2012athermal, bialke2013microscopic,buttinoni2013dynamical},  and  enhanced sedimentation~\cite{enculescu2011}.  
%
%% Moreover, the non-equilibrium character of the self-propulsion is even manifested in the absence of  particle interactions.   
%
%% For instance, the mean square displacement can display, unlike regular Brownian particles,  a complex behavior with several crossovers between ballistic and diffusive regimes, and a diffusion coefficient function of the active speed and tumbling frequency~\cite{peruani2007,  romanczuk2011}. 
%
%% 

%%% Fig 1 %%%%%%%%%%%%%%%%%%%%%%%%%%%%%%%%%%%%%%%%%%%%%%%%%%%%%%%%%
\begin{figure}[!]
\centering
%% \resizebox{\columnwidth}{!} {\includegraphics{fig1.eps}}
\resizebox{6cm}{!}{\rotatebox{0}{\includegraphics{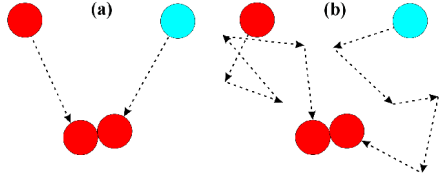}}}
\caption{The figure shows two scenarios [(a) aand (b)] where a red agent "activates" a blue one agent after an encounter/collision among the two agents. 
In  (a)  particles move in straight trajectories (ballistic regime) in between collisions, while in (b) particles perform several turns (diffusive regime).  
Note that in an over-damped dynamics, a collision event lasts a finite time. 
Importantly, the steady state densities of particles species (e.g. density of red and blue agents) depend on the motility parameters of the active particles, namely 
  active speed $v_0$ and tumbfling frequency $\lambda$. 
For movies, see SI. 
}  \label{fig:sketch}
\end{figure}
%%% Fig 1 %%%%%%%%%%%%%%%%%%%%%%%%%%%%%%%%%%%%%%%%%%%%%%%%%%%%%%%

Here, we aim at studying the spreading of information/signaling among actively moving units that do not display large-scale order.  
With this goal in mind, we analyze a system of self-propelled disks that perform run-and-tumble motion, where particles can adopt more than one internal state, see Fig. \ref{fig:sketch}. 
One of those internal states can be transmitted to other disk if the disk carrying this state maintains physical contact with another disk for a finite period of time. 
Making an analogy to chemical reaction, we refer to this process as a reaction process and say that  particles in the same internal state belong to the same particle species.  
The objective of current study  is to understand how the steady state densities of particle species depend on the motility parameters 
of the self-propelled disks -- shedding light  on the way information spreads in active systems -- in the context of reaction processes with an absorbing state. 
It is worth recalling that phase transitions into absorbing states is one of the fundamental problems in non-equilibrium 
statistical physics~\cite{hinrichsen2000non,krapivsky2010kinetic,marro}.  Here we intend to make a first step to understand how activity in the form of self-propulsion affects this classical problem of nonequilibrium statistical physics. 
It is important to stress that the understanding of the role of particle motion in reaction processes is key 
for a large number of applications beyond the context of active systems. For instance, it is well-known 
that  the average  concentration of (non self-propelled) chemical elements  depends on how the system is stirred~\cite{taylor09, tinsley09}. 
In  ecology,  the mobility of individuals  
affects the level of biodiversity~\cite{reichenbach07a, reichenbach07b}. 
In epidemics, it has been shown that the motion of individuals impacts the statistics of disease outbreaks~\cite{kuperman2001, boguna2003, colizza2007, hufnagel2004, gonzalez2004, miramontes2002}. 
In microbiology, the spreading of pathogens~\cite{weinbauer1998, beretta1998} and the spatial distribution of gene expressions and cell types~\cite{dworking,reichenbach07a} are known to be correlated to cell motility.  
Finally, in the context of social dynamics, it has been shown that  moving-agent models   can be used 
as a proxy to mimic realistic social dynamics~\cite{gonzalez2006}, 
which paved the way to study opinion dynamics in moving-agent systems~\cite{terranova2014, clementi2015}. 
In summary, studying how the spreading of information/signaling is affected by the motility 
parameters of active particles is a fundamental problem in active matter -- and in the broad context of nonequilirbium statistical mechanics -- which may have important implications for a large number of applications.

%Here, we unveil the role of the active speed and tumbling frequency  on the  statistics 
%of a simple chemical reaction, that exhibits  an absorbing and an active phase. The reaction occurs among self-propelled disks that perform a run-and-tumble motion and interact by volume exclusion, see Fig. \ref{fig:sketch}. 
%%
%We find that the average density of the chemical species exhibits several regimes with active speed and tumbling frequency, and compute for a set of reaction rates, 
%the critical velocity and tumbling rate that allows the system to stay in the active phase.  

%%%%%%%%%%%%%%%%%%%%%%%%%%%%% Particle motion %%%%%%%%%%%%%%%%%%%%%%%%

\section{Model definition} 
\subsection{Particle motion}
We consider, as in~\cite{peruani2008}, a two-dimensional system of $N$ self-propelled disks moving in a box of linear size $L$ with periodic boundary conditions.  
The equation of motion of the 
$i$-th disk is given by: 
\begin{eqnarray} 
\label{update_position} \dot{\mathbf{x}}_i(t) &=& 
v_0 \mathbf{v}(\theta_i)  -  \sum_{j\neq i} 
{\mathbf{\nabla}U(\mathbf{x}_i(t), \mathbf{x}_j(t))} \, ,
\end{eqnarray} 
where $\mathbf{x}_i(t)$ is the position of the particle, $v_0$ is the active speed, $\mathbf{v}(\theta_i) =
(\cos(\theta_i),\sin(\theta_i))$ with  $\theta_i$ an angle that denotes the  propulsion direction. 
The propulsion direction $\theta_i$ obeys a classical Poisson process: at rate $\lambda$ -- which we refer to as tumbling frequency -- 
a new angle is selected from the interval $[0, 2\pi)$. 
This defines a classical run-and-tumble  process~\cite{berg1993random,cates2013active},  whose  distribution $p(\theta_i,t)$ obeys  $\partial_t p(\theta_i,t)= - \lambda p(\theta_i,t) + \int \, d\theta^\prime \, T(\theta^\prime \to \theta_i)p(\theta^\prime, t)$, where $T(\theta^\prime \to \theta_i) = \frac{\lambda}{2\pi}$ is the transition probability  from $\theta^\prime \to \theta_i$. Note that $p(\theta_i,t\to\infty) = \frac{1}{2\pi}$ and the particle stays in a given direction  a characteristic time $\lambda^{-1}$.    
In between turnings, particles interact through a soft-core potential $U$,  which penalizes particle  
overlapping. Specifically, we implement a two-body repulsive potential, 
which depends on the distance between the center of mass of the two disks as follows: 
\begin{equation} 
\label{twobody_potential} U(\mathbf{x}, \mathbf{x'}) = \left\{ 
\begin{array}{ll} 
{a(v_0)} \left[ \left(\frac{|\mathbf{x} - \mathbf{x'}|}{2r}\right)^{- b} - 1 \right] & \mbox{if $|\mathbf{x} - \mathbf{x'}|<2r$}\\ 
\\ 
0 & \mbox{if $|\mathbf{x} - \mathbf{x'}| \geq 2r$}\\ 
\end{array} \right. 
\end{equation} \\ 
where $r$ is the radius of the disks, $b$ is a constant, and 
$a(v_0)$ is a linear function of $v_0$, $a(v_0) = c_1 + c_0 v_0$ such that the maximum 
overlapping area between two agents is independent of 
$v_0$. By appropriately choosing the units of $a$, we have absorbed the mobility constant in such that  $\mathbf{\nabla}U$ has units of speed instead of force. 
Note that $b$ controls how soft/hard is the potential $U$, becoming increasingly harder as $b$ is increased. 
In the following we fix the parameters $r=1$, $c_0=1.62$ (in units of distance), $c_1=10^{-4}$ (in units of speed squared), $b=1$, 
and the density to $d_0=N/L^2=0.045$ ensuring that for the explored range of $v_0$ and $\lambda$ the system remains in the gas phase.    
%% In absence of interactions and neglecting speed fluctuations, 
Under these conditions, the motion of the self-propelled disks  can be approximated by  F{\"u}rth's formula:  
\begin{eqnarray} 
\label{eq:msq_displacement} \langle \mathbf{x}^2(t) \rangle & \simeq & 
2 \frac{ v_0^2}{\lambda^2} (\lambda t - 1 + e^{-\lambda t})  \,, 
\end{eqnarray} 
and thus, for $t<<\lambda^{-1}$ we can consider that particles move
ballistically at speed $v_0$, while for $t>>\lambda^{-1}$ their motion is
characterized by a diffusion coefficient  $D = v_0^2 \lambda^{-1} $.  
%

%%%%%%%%%%%%%%%%%%%%%%%%%%%%% Chemical Reaction %%%%%%%%%%%%%%%%%%%%%%

\subsection{Reaction processes} 
Our intention is to make a necessary step towards an understanding of phase transitions into absorbing states by 
studying the dependencies of the equilibrium densities of particles species with  motility parameters. 
Given our goal, any reaction process with an absorbing state serves to our purpose. 
Prototypical examples of such reaction processes are -- using the terminology of epidemics~\cite{murray} -- 
the Susceptible-Infected-Susceptible (SIS) reaction, which defines the so-called contact process in physics~\cite{hinrichsen2000non,krapivsky2010kinetic,marro}  
or the Susceptible-Infected-Recovered-Susceptible (SIRS) dynamics, which defines a simple spatially extended excitable system, e.g. the Forest-Fire model~\cite{hinrichsen2000non,krapivsky2010kinetic}. 
Note that though here we use terminology of epidemics, many physical systems fall in the same universality class as indicated in~\cite{hinrichsen2000non}.
Given that for dilute systems, we expect the SIRS model to behave as the SIS model for fast $R \to S$ transitions, and to recover the SIR model in the absence of such transition. Thus, to remain as general as possible, we choose 
the SIRS model, whose dynamics is defined by:
\begin{eqnarray}\label{eq:reactions}
\begin{array}{ccc}
S+I \stackrel{\alpha}{\rightarrow}  2I , & 
I  \stackrel{\beta}{\rightarrow}       R , &
R \stackrel{\gamma}{\rightarrow} S \, ,
\end{array}
\end{eqnarray}
where $\alpha$, $\beta$, and $\gamma$ are (constant) transition rates. Note that 
the reaction $S+I\to2I$ to occur requires a particle of the species ``S" and an particle of the species ``I" to maintain physical contact for a finite time. 
We say that two particles are in physical contact whenever the center of mass of these two particles, e.g. $\mathbf{x}$ and $\mathbf{x}^\prime$ are such that $U(\mathbf{x}, \mathbf{x}^\prime)>0$. 
Note that since we are considering active particles that obey an over-damped dynamics, see Eq.~(\ref{update_position}), collisions are not instantaneous and last a finite time; and more  importantly, during a collision particles maintain physical contact.  
In the following,  without loss of generality we fix $\beta^{-1} = 200$ and $\gamma^{-1}=500$ (in arbitrary time units) and vary the reaction rate $\alpha$ since it is the only  rate connected to a transition that is affected by particle motion.

%%% Fig 2 %%%%%%%%%%%%%%%%%%%%%%%%%%%%%%%%%%%%%%%%%%%%%%%%%%%%%%%%%
\begin{figure}
\centering
%% \resizebox{\columnwidth}{!} {\includegraphics{fig2.eps}}
\resizebox{7cm}{!}{\rotatebox{0}{\includegraphics{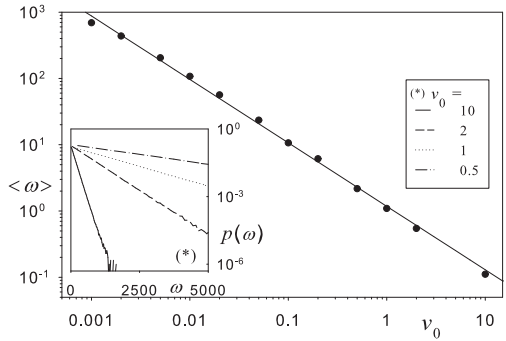}}}
\caption{Collision statistics. Average collision duration time $\langle \omega \rangle$ as function of the active speed $v_0$. 
Symbols correspond to simulations, while the solid line is a power-law fit, see Eq.~(\ref{eq:scaling_omega}).  
The inset  shows that $p(\omega)$ is exponential distributed. The different curves correspond to various values of $v_0$. 
}  \label{fig:omega}
\end{figure}
%%% Fig 2 %%%%%%%%%%%%%%%%%%%%%%%%%%%%%%%%%%%%%%%%%%%%%%%%%%%%%%%

\section{Mean-field}
At higher densities and active speeds, a system of self-propelled disks can undergo 
a phase separation~\cite{fily2012athermal,bialke2013microscopic}. 
However, here we use parameters that ensure that the 
system  remains in a gas-like phase, where clusters are small, collisions are mainly binary, 
and the system is well-mixed. 
Under these conditions, the temporal evolution of the densities of particle species S, I and R 
can be described by a mean-field approach of the form: 
\begin{eqnarray}
\label{low_s} \dot{\rho_S} &=& \gamma (1-\rho_S-\rho_I) - \psi(v_0, \lambda, \alpha)
R(d_0, v_0, \lambda) \rho_I \rho_S \, , \\
\label{low_i} \dot{\rho_I} &=& \psi(v_0, \lambda, \alpha) R(d_0, v_0, \lambda) \rho_I
\rho_S - \beta \rho_I \,, 
\end{eqnarray}
where the dot denotes time derivative and $\rho_S$, $\rho_I$, and
$\rho_R$ are normalized densities such that $\rho_S+\rho_I+\rho_R=1$; the actual densities correspond to $\rho_k \, d_0$, with $k \in \{S, I, R \}$ and $d_0 = N/L^2$  the (global) density of particles. 
Note that in deriving Eqs. (\ref{low_s}) and (\ref{low_i}) we have used that due to particle conservation $\rho_R = 1 - (\rho_S+\rho_I)$. 
For details on the derivation of population mean-field models, see e.g.~\cite{murray}.  
It is important to understand that in a standard all-to-all mean-field scheme the expected term 
in front of  $\rho_I \rho_S$ in Eqs. (\ref{low_s}) and (\ref{low_i}) is $\alpha$ times a constant (i.e. $N$ for $N \gg 1$). 
However, here due to the active motion of the disks, particles are rarely in contact and when they do it, it is only 
for a short period of time. 
The question we pose is then how to effectively renormalize the transition rate $I \to S$ and how such an effective transition rate depends on the motility parameters.  
To compute an effective transition rate, we need to estimate the rate at which particles meet -- let us call this rate the collision frequency and denote it by  $R(d_0, v_0, \lambda)$  -- 
and the probability that a reaction $I \to S$ occurs for a random encounter between a particle S and a particle I; let us use the symbol $\psi(v_0, \lambda, \alpha)$ to refer to this probability. 
In summary, $R(d_0, v_0, \lambda)\psi(v_0, \lambda, \alpha)$ defines the effective transition rate $I \to S$. Now, let us assume we know $R$ and $\psi$ (we estimate both of them below). 
It is straightforward to verify that 
Eqs.~(\ref{low_s}) and~(\ref{low_i}) have two steady states.  One of these states,  
often referred to as the {\it absorbing phase}, is given by $\rho_{S}(t\to\infty)=1$ and
$\rho_{I}(t\to\infty)=0$. The other one, called the {\it active phase}, is
given by:  
\begin{eqnarray}
\label{low_s_st} \rho_{S}(t\to\infty) &=& \rho_{S}^{*} =  \frac{\beta}{\psi(v_0,
  \lambda, \alpha)\,R(d_0,v_0,\lambda)} \, ,\\ 
\label{low_i_st} \rho_{I}(t\to\infty) &=& \rho_{I}^{*} =  \frac{\gamma}{\gamma+\beta} \left(1-\rho_{S}(t\to\infty) \right) \,.
\end{eqnarray}
Strictly speaking, Eqs.~(\ref{low_s_st}) and~(\ref{low_i_st}) correspond  to the (active) steady states of an infinite system of density $d_0$. Finite size fluctuations, which we have neglected here, may lead to deviation of mean-field predictions.  
The linear stability analysis around the absorbing state ($\rho_S=1$, $\rho_I=0$) -- obtained by inserting $\rho_S = 1 - \delta(t)$ and $\rho_I = \delta(t)$ into Eq.~(\ref{low_i_st}) and keeping linear terms in $\delta$ -- 
 provides the following condition (within the mean-field approximation) for the existence of the active phase:  
\begin{eqnarray}
\label{eq:threshold_cond} 
\psi(v_0, \lambda, \alpha) R(d_0, v_0, \lambda)  > \beta  \, .
\end{eqnarray}
In the following, we provide estimates for $R(d_0, v_0, \lambda)$ and $\psi(v_0, \lambda, \alpha)$.

%%% Fig 1 %%%%%%%%%%%%%%%%%%%%%%%%%%%%%%%%%%%%%%%%%%%%%%%%%%%%%%%%%
\begin{figure}
\centering
\resizebox{\columnwidth}{!} {\includegraphics{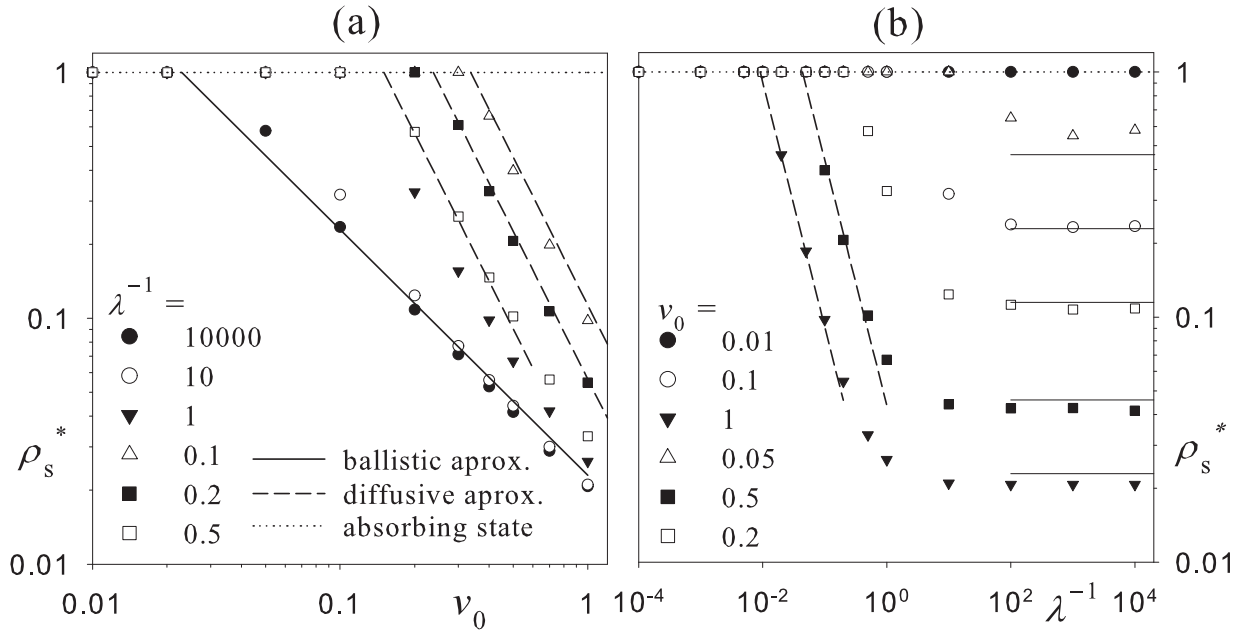}}
\caption{Instantaneous transmission. (a) $\rho_S^{*}$ vs. $v_0$ for various values of $\lambda$. 
(b) $\rho_S^{*}$ vs. $\lambda^{-1}$ for  various values of $v_0$. 
In (a) and (b), symbols correspond to simulations, while solid lines  to the ballistic  and dashed line to the diffusive approximation. 
The absorbing state is indicated by the horizontal dotted line.
}  \label{fig:rho_notrans}
\end{figure}
%%% Fig 1 %%%%%%%%%%%%%%%%%%%%%%%%%%%%%%%%%%%%%%%%%%%%%%%%%%%%%%%
\subsection{Time in between collisions -- estimating R} 
To estimate $R(d_0, v_0, \lambda)$, we consider that there are two clearly distinct regimes  
that we can identify by constructing a dimensionless quantity ($l_{ball}/l_{\rho}$) that results from the ratio between two characteristic length scales in the system:  
 the  characteristic distance $l_{\rho}$ in between particles --  neglecting clustering effects and
assuming a homogeneous distribution of particles -- which is given by
$l_{\rho}=1/\sqrt{d_0}$, and  the typical distance $l_{ball}$ that particles move in straight line (i.e. the typical distance in between two tumbling events).  
Since the typical time in between tumbling events is
$\lambda^{-1}$, then $l_{ball}= v_0 \lambda^{-1}$. 
Now, we are in condition of defining the {\it ballistic} and {\it diffusive} spreading regimes. 
We call ballistic spreading, the regime in which we can 
ensure that in between collisions, active particles  move  ballistically, i.e. in straight lines.  
The condition for this regime is given by:
\begin{eqnarray}
\label{BallisticCoef}
 l_{ball}/l_{\rho} = v_0 \lambda^{-1} \sqrt{d_0} \gg 1 \,.
\end{eqnarray}
Under this condition, we can make use of kinetic gas theory~\cite{reif2009fundamentals} and approximate $R(d_0, v_0, \lambda)$  as:
\begin{eqnarray}
\label{R_ballistic}
R(\rho, v_0, \lambda) \sim v_0 \sigma_0 d_0 \,,
\end{eqnarray}
where $\sigma_0=4r$ is the scattering cross section of particles ($v_0$ and $d_0$ were defined above).
The diffusive spreading regime corresponds to the opposite limite:  $l_{ball}/l_{\rho}\ll1$.  
In this regime,  active particles perform several tumbling events in between collisions.  
This implies that the time in between collisions is much larger than $\lambda^{-1}$, 
which according to Eq.~(\ref{eq:msq_displacement}) means that the active particles are deep inside the diffusive regime.  
In other words, the time in between collisions is dictated by a diffusive process characterized by a diffusion constant $D$, whose expression was given above after Eq.~(\ref{eq:msq_displacement}). 
To estimate the time in between collisions -- let us refer to it as $\tau$ -- we consider that the centers of mass of two neighboring self-propelled disks are separated an average distance 
$1/\sqrt{d_0}$ and that the average distance the disks have to travel to collide is $1/\sqrt{d_0}-2r$, since the disks have a radius $r$. Then,  
under the assumption we are in the diffusive regime, we expect $\tau$ to be such that  $\left( 1/\sqrt{d_0} -2r \right)^2 \simeq D \tau$. 
Thus, the collision frequency $R$, which is the inverse of $\tau$, takes the form:
\begin{eqnarray}
\label{R_diffusive}
R(d_0, v_0, \lambda) \sim v_0^2 \lambda^{-1}  \left(\frac{1}{\sqrt{ d_0}} -2r \right)^{-2}  \,.
\end{eqnarray}
Note that the above expression is only an approximation. A rigorous calculation of the collision frequency would require to solve a first passage time problem; for details see~\cite{redner2001guide, krapivsky2010kinetic}.

%%% Fig 2 %%%%%%%%%%%%%%%%%%%%%%%%%%%%%%%%%%%%%%%%%%%%%%%%%%%%%%%%%
\begin{figure}
\centering
\resizebox{\columnwidth}{!} {\includegraphics{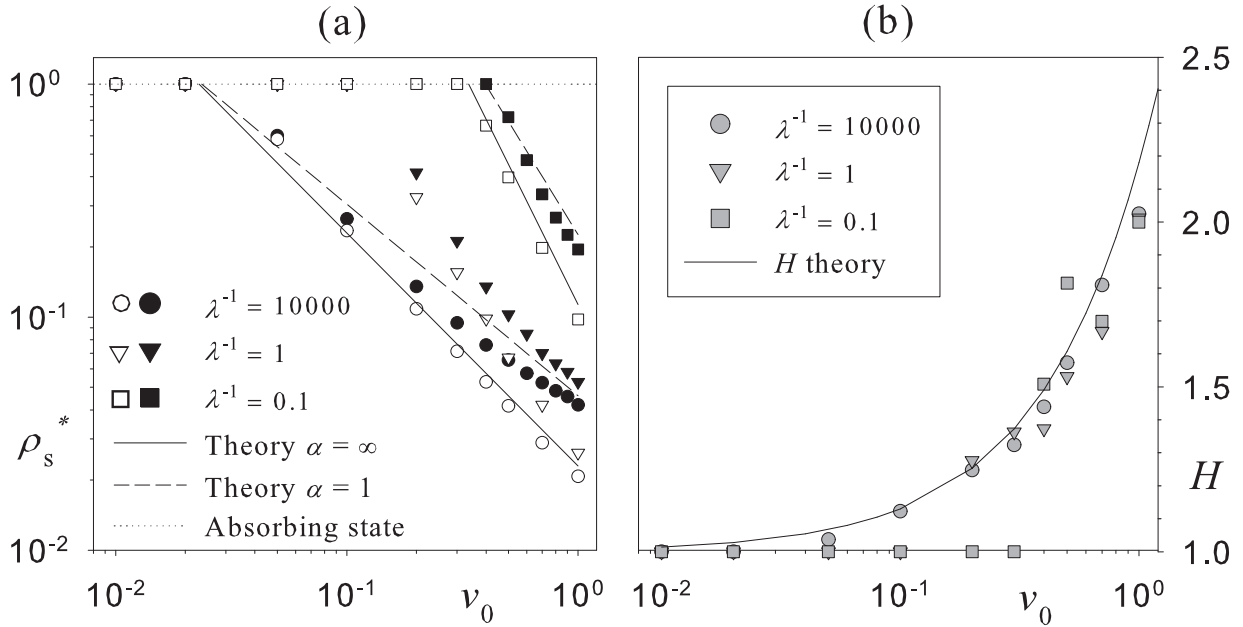}}
\caption{Finite transmission - (a) $\rho_S(t \to \infty)$ vs. $v_0$ for various $\lambda$ values. 
Open symbols correspond to simulations with $\alpha \to \infty$, while solid symbols to $\alpha=1$. 
The solid and dashed curve correspond to the mean-field approximation for   $\alpha \to \infty$ and $\alpha=1$, respectively, in both ballistic and diffusive approximation. 
(b) $H=\rho^{*}_S(\alpha)/\rho^{*}_S(\infty)$ vs. $v_0$ for various $\lambda$ values. 
Simulations (symbols) follow the mean-field prediction (solid curve) as soon as the system moves into the active phase. 
}  \label{fig:rho_trans}
\end{figure}
%%% Fig 2 %%%%%%%%%%%%%%%%%%%%%%%%%%%%%%%%%%%%%%%%%%%%%%%%%%%%%%%

\subsection{Collision duration -- estimating $\psi$} 
A collision between two particles is a relatively slow 
process in which particles stay in contact -- understanding by this that the potential energy $U$ between the two particles is larger than $0$ -- for a finite time $\omega$. 
By looking at histograms of  $\omega$ in simulations, we learn that  $\omega$ is exponentially distributed -- meaning that $p(\omega)=\exp(-\omega/\langle \omega \rangle)/\langle \omega \rangle$ --, see inset in Fig.~\ref{fig:omega}.  
Furthermore, by performing a systematic study of $\langle \omega \rangle$ vs. $v_0$, Fig.~\ref{fig:omega}, we find that $\langle \omega \rangle$ follows  a power-law 
with $v_0$: 
\begin{eqnarray}
\label{eq:scaling_omega}
\langle \omega \rangle = k\, v_0^{-\xi} \, ,
\end{eqnarray}
with $k=1.18$ and $\xi=0.957$ for $\lambda^{-1}$ larger than $r/v_0$~\footnote{For $\lambda^{-1}$ smaller or equal to $r/v_0$, we find that $\xi \sim 0.5$. This regime is out of the scope of the current study and will be analyzed elsewhere.}.  
Knowing the statistics of $\omega$, now we can focus on obtaining an estimate for the probability 
that, during a collision event of duration $\omega$ between a particle I and a particle S, a reaction $S+I\to 2I$ occurs. 
 We make use of the fact that the reaction is given by a simple  Poissonian process~\cite{van1992stochastic}, 
 which let us compute the probability that a reaction takes place in the time interval $\omega$ as $1-e^{-\omega \alpha}$, 
 under the  assumption that the initial particle I does not transition to R in this time interval. 
 The next step to estimate $\psi$ is to make an average over all possible collision durations $\omega$ -- that we know is distributed exponentially as indicated above -- 
 to  express $\psi$ as: 
\begin{eqnarray}
\label{collision_duration}
\psi(v_0, \lambda, \alpha) = \int_0^{\infty} d\omega\, p(\omega) \, \left( 1 - e^{-\omega \alpha}
\right) = \frac{\alpha \langle \omega \rangle}{1+\alpha \langle \omega \rangle} \, .
 %% \sim  1 - e^{- \langle \omega  \rangle \alpha}  \,
%
\end{eqnarray}
%
%% where we have used the fact that the $p(\omega)$ is an exponential distribution: $p(\omega)=\exp(-\omega/\langle \omega \rangle)/\langle \omega \rangle$; see inset in Fig.~\ref{fig:omega}. 
%
%
%
Thus, inserting Eq.~(\ref{eq:scaling_omega}) into
Eq.~(\ref{collision_duration}), we find that $\psi = (1+ k^{-1} \alpha^{-1} v_0^{\xi})^{-1}$.

\section{Comparison with simulations} 
%
% {\color{red} We performed numerical simulations to study the evolution of the model described above. 
%The steady state values are obtained by performing an average of the populations values observed after a transient time. 
%To perform all the simulations we choose the set of parameter values (in arbitrary units) given in ~\footnote{
%%
%Simulation parameters. Parameters associated with agent motion:
% $a/\zeta= 3.24 \upsilon $, $b=1$, and $r=1$.
%%% $\gamma/\zeta=0.9506$
%%%
%Associated to disease dynamics: $1/\beta = 200$, $1/\gamma = 500$. 
%%
%Most simulations were carried out using  $d_0 = 0.045$.
%%
%%% Corresponding to the fitting of $\omega$: $K=1.042$ and $\xi = 0.916$.
%%
%The values of $\upsilon$, $\lambda$ and $\alpha$ are specified in the text.} . }
% 
Let us start by analyzing an infinitely fast reaction rate $\alpha$, which let us take the limit $\alpha \to \infty$ in Eq.~(\ref{collision_duration}) and express  $\psi=1$. 
The prediction is that for $l_{ball}/l_{\rho}\gg 1$, by inserting Eq.~(\ref{R_ballistic}) into
Eq.~(\ref{low_s_st}), we find $\rho_S^{*} \sim v_0^{-1}$, solid curve in Fig.~\ref{fig:rho_notrans}(a), while  
 by inserting Eq.~(\ref{R_diffusive}) into
Eq.~(\ref{low_s_st}), we obtain $\rho_S^{*} \sim \lambda\,v_0^{-2}$, dashed curves in Fig.~\ref{fig:rho_notrans}(a). 
In summary, there are at least two scalings of $\rho^{*}_S$ with  $v_0$, 
and a crossover between these two scalings at $l_{ball}/l_{\rho} \simeq 1$. 
%
%% Fig.~\ref{fig:rho_notrans}(a) confirms that in simulations the scaling with $v_0$ is  as mentioned above. 
%
%%% Note that for intermediate values of $\lambda^{-1}$, the same curve $\rho^{*}_{S}$ can exhibit an initial scaling $\propto v_0^{-2}$ for low values of $v_0$, followed by  the ballistic scaling  $\propto v_0^{-1}$ for large $v_0$ values. 
%
%
Fixing $v_0$, the set of mean-field approximations detailed above predicts that 
at large enough tumbling frequencies $\lambda$ such that $l_{ball}/l_{\rho}\ll 1$, $\rho_S^{*} \propto \lambda$, dashed curves in Fig.~\ref{fig:rho_notrans}(b). As $\lambda$ is decreased,  the system should cross 
$l_{ball}/l_{\rho}=1$ and $\rho_S^{*}$ becomes independent of $\lambda$, as confirmed by the horizontal solid lines in Fig.~\ref{fig:rho_notrans}(b). 

Now, we move to finite reaction rates  $\alpha$,  where the approximations given above suggest that  the spreading dynamics is strongly affected by the average collision duration $\langle \omega \rangle$ through Eq.~(\ref{collision_duration}). 
The predictions  are now $\rho_S^{*} \sim v_0^{\xi-1}$ for the ballistic regime, that consists 
of inserting Eq.~(\ref{R_ballistic}) and Eq.~(\ref{collision_duration}) into Eq.~(\ref{low_s_st}),   
and $\rho_S^{*} \sim \lambda\,v_0^{\xi-2}$ for the diffusive regime, obtained by inserting Eq.~(\ref{R_diffusive}) and Eq.~(\ref{collision_duration}) into Eq.~(\ref{low_s_st})~\footnote{In both cases, the factorization was performed assuming $\alpha k<v_0^{\xi}$.}. 
Fig. \ref{fig:rho_trans}(a) shows a direct comparison between the prediction for finite (dashed curves) and infinitely fast reaction rates $\alpha$ (solid curves). 
%
%% Deviation between finite and infinite reaction rates are due mainly (but not exclusively) to the statistics of the collision duration. 
%
In order to get a direct understanding of the role of the probability $\psi$, we plot in Fig. \ref{fig:rho_trans}(b) $H = \rho_S^{*}(\alpha)/ \rho_S^{*}(\alpha \to \infty)$ (let us recall that   $\rho_S^{*}(\alpha\to\infty) \sim v_0^{-1}$ 
and $\rho_S^{*}(\alpha \to \infty) \sim \lambda\,v_0^{-2}$ for the ballistic and diffusive regime, respectively). 
Thus, according to the approximations developed in the previous sections, for both regimes, i.e. ballistic and diffusive, $H \sim (1+ k^{-1} \, \alpha^{-1} \, v_0^{\xi})$, see solid curve in  Fig. \ref{fig:rho_trans}(b).

\begin{figure}
\centering
\resizebox{7.2cm}{!}{\rotatebox{0}{\includegraphics{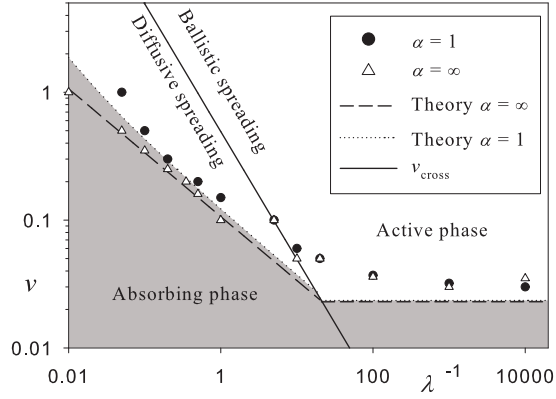}}}
%% \resizebox{\columnwidth}{!} {\includegraphics{phaseDiagram.eps}}
\caption{Phase diagram $v$ vs $\lambda^{-1}$. Symbols 
correspond to the absorbing-active boundary as  found in simulations. 
Dashed and dotted lines correspond to the critical point ($\lambda_c^{-1}$, $v_c$) predicted by the 
mean-field approximation for $\alpha \to \infty$ and $\alpha =1$, respectively. 
The solid line, $v_{cross}$, indicates the crossover between ballistic and diffusive spreading. 
}  \label{fig:phaseDiagram}
\end{figure}

In the following, we look for the critical active speed $v_c$ (and tumbling rate $\lambda_c$) above (below)  which the active phase should be observed. 
We are particularly interested in knowing, given a set of parameters $\alpha$, $\beta$, $\gamma$, and $d_0$, the behavior of $v_c$ with the tumbling rate $\lambda$. 
Inserting  Eq.~(\ref{R_ballistic}) and  Eq.~(\ref{collision_duration}) into Eq.~(\ref{low_s_st}), and by requesting $\rho_S^{*} =1$, we derive a condition for $v_c$ which does not depend on $\lambda$;  see horizontal lines in Fig.\ref{fig:phaseDiagram}. 
Similarly, by inserting  Eq.~(\ref{R_diffusive}) and  Eq.~(\ref{collision_duration}) into Eq.~(\ref{low_s_st}), and under the same condition,  we find that $v_c$ is solution of 
$v_c^2 - q \lambda \epsilon v_c^{\xi}= q \lambda$, where $q=\beta (1/\sqrt{ d_0}-2r)^2$ and $\epsilon=1/\alpha$. 
In order to obtain a close expression for $v_c$, we  assume that $\epsilon$ is small and $\xi = 1 + \delta$, with $| \delta | \ll 1$, which leads to:  
\begin{eqnarray}
\label{eq:vc_diff}
%
%% v_c = \frac{\epsilon q \lambda}{2} + \sqrt{q \lambda} \sqrt{1+ \frac{\epsilon^2 q \lambda}{4}}\, , 
v_c = \frac{\epsilon q \lambda}{2} + \left[ q \lambda \left(1+ \frac{\epsilon^2 q \lambda}{4} \right) \right]^{-1/2}
%% \sqrt{q \lambda} \sqrt{1+ \frac{\epsilon^2 q \lambda}{4}}\, , 
%
\end{eqnarray}
where  terms proportional to $\epsilon \delta$  have been neglected.
For infinitely fast reactions, $\epsilon \to 0$, and $v_c \propto \lambda^{1/2}$. Note that this  result  is independent of $\delta$.  
For large, but finite reaction rates, Eq.~(\ref{eq:vc_diff}) provides a good estimate of $v_c$ as shown in Fig. \ref{fig:phaseDiagram}.  
Using Eq.~(\ref{BallisticCoef}) and the fact that the ballistic and diffusive approximation for $\rho_S^{*}$ coincide at the crossover point, we find  $v_{cross}(\lambda_c)$, the solid curve  in Fig. \ref{fig:phaseDiagram}. 
%
%% FP - Though beyond the scope of the present work, the MF prediction for $\alpha << 1$ is $v_c \propto \lambda^{1/{2-\xi}}$, a value at which Eq.~(\ref{eq:vc_diff}) tends as $\epsilon$ get larger.  

%%% Fig 3 %%%%%%%%%%%%%%%%%%%%%%%%%%%%%%%%%%%%%%%%%%%%%%%%%%%%%%%%%
%%% Fig 3 %%%%%%%%%%%%%%%%%%%%%%%%%%%%%%%%%%%%%%%%%%%%%%%%%%%%%%%

\section{Conclusions} 
We have studied, combining individual-based simulations and mean-field arguments,  the dependency of the equilibrium densities of particle species with motility parameters 
in an active system consisting of self-propelled disks. Importantly, we consider a reaction process with an absorbing and active state. 
For a finite reaction rate $\alpha$,  we found that there are two distinct regimes in the active phase with the active speed $v_0$ and tumbling frequency $\lambda$ 
that lead to  steady state densities $\rho_S^* \propto v_0^{\xi-1}$ and $ \rho_S^* \propto \lambda v_0^{\xi-2}$ for the ballistic and diffusive regime, respectively. 
%
%% The crossover between these two regimes is given by a simple condition,  Eq.~(\ref{BallisticCoef}), that depends on $d_0$, $v_0$, and  $\lambda$. 
%
For a given set of reaction rates $\alpha$,  $\beta$, and $\gamma$, and  density $d_0$, we have been able to compute the phase-diagram in terms of motility parameters, namely  
$v_c$ and $\lambda_c$, that separates the absorbing and active phase. 
These results were obtained from a combination of stochastic simulations of the microscopic dynamics and mean-field arguments. 
It is important to stress that the developed mean-field arguments include information on the spatial dynamics of the self-propelled disks. 
Moreover, the results here derived cannot be obtained by a classical reaction-diffusion process as done in~\cite{murray}, since these approaches decouple particle motility and effective reaction rates and assume that 
particle transport is always diffusive.  
A word of warning: the here derived arguments neglect spatial correlations among the particle species.  
At higher densities,  lower dimensions, and/or close to the critical point, deviations between the derived mean-field arguments and simulations are expected,  
 and
 correlation and fluctuations  should be taken into account  in order to obtain 
an accurate description of the system dynamics along the lines explained in~\cite{lee2013} 
that makes used of the formalism developed in~\cite{peliti1985}.  

Here, we have shown that  the (transient) behavior of the active particles in between collisions plays a key role to understand the equilibrium densities of particle species. 
While here we considered only ballistic and diffusive regimes, we can imagine a more general scenario where  $\langle \mathbf{x}^2 \rangle \propto t^{\chi}$ with $1\leq \chi \leq 2$ 
that we expect to lead to different scalings of the equilibrium densities with  active speed $v_0$ and tumbling frequency $\lambda$.  
Importantly, the presence of different transport regimes in systems of active particles suggests that phase transitions into an absorbing state 
do not necessarily fall into the directed percolation class or related universality classes for active systems~\cite{hinrichsen2000non, marro}. 
Understanding in which universality class fall ``reactive" active systems with an absorbing state remains a fundamental, challenging question that 
can only be addressed by combining large-scale simulations and renormalization group techniques, which we hope will be the subject of future works.  

The simple, but fundamental results reported in this study represent a necessary first step to a better understanding on the spreading of information/signaling among actively moving units, 
an issue of key importance in the context of reaction processes and diseases~\cite{weinbauer1998, beretta1998, dworking, reichenbach07a}, synchronization among moving oscillators~\cite{skurfa04, frasca08, peruani2010b, fujiwara2011, grossmann2016}, 
 organization of self-propelled particles into collective motion~\cite{vicsek1995novel, toner1995long, ginelli2010large, barberis2016, vicsek2012, marchetti2013},  as well as several technological and biomedical applications involving moving entities~\cite{woodhouse2017, din2016}.  

\section*{Acknowledgements}
F. P. was supported by the Agence Nationale de la
Recherche via project BactPhys, Grant No. ANR-15-
CE30-0002-01. 

%%% ~\cite{flocking}. 
%% Review Marchetti + Vicsek + barberis
%% Antes citar estos no olvidar en moving oscillator uno de Pagonabarraga + ver si Botinelli tambien se añade
%
%% 

%%%%%%%%%%%%%%%%%%%%%%%%%%%% Bibliography %%%%%%%%%%%%%%%%%%%%%%%%%%%%%%%%%%%%%%%%

%%% \bibliography{biblio_std_chemicalreact}

%%%%%%%%%%%%%%%%%%%%%%%%%%%%%%%%%%%%%%%%%%%%%%%%%%%%%%%%%%%%%%

%%%%%%%%%%%%%%%%%%%%%%%%%%%%%%%%%% Figures %%%%%%%%%%%%%%%%%%%%%%%%%%%%%%%%%%%%%%%%%%%%

\end{document}